\documentclass[twocolumn,amssymb, nobibnotes, floatfix, aps, prb]{revtex4-2}
\usepackage[bookmarks=false]{hyperref}
\usepackage[utf8]{inputenc}
\usepackage[english]{babel}
\usepackage{amssymb}
\usepackage{amsmath}
\usepackage{graphicx}
\usepackage{color}
\usepackage{physics}
\usepackage{float}
\renewcommand{\vec}[1]{\boldsymbol{#1}}
\graphicspath{{Figs/}}
\newcommand{\be}{\begin{equation}}
\newcommand{\ee}{\end{equation}}
\newcommand{\bea}{\begin{eqnarray}}
\newcommand{\eea}{\end{eqnarray}}

\def\nn{\nonumber}
\def\lb{\label}
\def\pref{\eqref}

\def\d{\delta}
\def\g{\gamma}
\def\m{\mu}

\def\th{\theta}

\def\tt{\hat \tau}

\begin{document}


\title{Electronic band structure and pinning of Fermi energy to van Hove singularities in twisted bilayer graphene: a self consistent approach
}

\author{Tommaso Cea$^1$}
\email{tommaso.cea@imdea.org}
\author{Niels R. Walet$^2$}
\email{Niels.Walet@manchester.ac.uk}
\homepage[]{http://bit.ly/nielswalet}
\author{Francisco Guinea$^{1,2}$}
\email{francisco.guinea@manchester.ac.uk}
\affiliation{$^1$Imdea Nanoscience, Faraday 9, 28015 Madrid, Spain}
\affiliation{$^2$School of Physics and Astronomy, University of Manchester, Manchester, M13 9PY, UK}

\date{\today}
\begin{abstract}
The emergence of flat bands in twisted bilayer graphene leads to an enhancement of interaction effects, and thus to insulating and superconducting phases at low temperatures, even though the exact mechanism is still widely debated. The position and splitting of the flat bands is also very sensitive to the residual interactions.
Moreover, the low energy bands of twisted graphene bilayers show a rich structure of singularities in the density of states, van Hove singularities, which can enhance further the role of interactions. We study the effect of the long-range interactions on the band structure and the van Hove singularities of the low energy bands of twisted graphene bilayers. Reasonable values of the long-range electrostatic interaction lead to a band dispersion with a significant dependence on the filling. The change of the shape and position of the bands with electronic filling implies that the van Hove singularities remain close to the Fermi energy for a broad range of fillings. This result can be described as an effective pinning of the Fermi energy at the singularity.
The sensitivity of the band structure to screening by the environment may open new ways of manipulating the system.

\end{abstract}

\pacs{???}

\maketitle
\section{Introduction}
The recent observation of
insulating behavior at charge neutrality (CN) and unconventional superconductivity close to CN
in twisted graphene bilayer (TBG) at the magic angle $\theta\sim 1^\circ$ \cite{cao_nat18,cao_nat18_2},
see also \cite{Ketal17,huang_prl18,Yetal19},
has motivated a renewed interest in the physics of twisted graphene superlattices and related structures.

Even more recent experiments have revealed unexpected features of magic-angle TBG close to CN,
such as the presence large spectral gaps at fractional filling \cite{tomarken_cm19}, the appearance of alternating superconducting and insulating phases on an increase of 
the occupation of the conduction bands \cite{efetov_cm19}
as well as evidence of charge ordering and broken rotational symmetry \cite{jiang_cm19,choi_cm19}.

From a theoretical point of view, the low energy continuum model of TBG originally proposed in \cite{LPN07}
and developed in \cite{BM11,SGG12,LPN12,KYKOKF18} leads to isolated flat bands at CN,
with vanishing Fermi velocity 
at the certain angles, the ``magic angles".
In contrast to tight-binding models \cite{shallcross_prb10,moon_prb13,lin_prb18},
which require a substantial numerical effort because of the large size of
the moir\'e superlattices at small twist angles  (with more than $10^4$ atoms per supercell), such 
continuum models are a much more efficient alternative. 
They have recently been generalized
to describe the electronic spectrum of related twisted materials,
such as  twisted graphene tri-layer \cite{amorim_cm19,mora_cm19},
quadruple-layer \cite{zhang_prb19,chebrolu_cm19,koshino_cm19}
and multi-layer graphitic structures \cite{khalaf_cm19,cea_cm19,liu_cm19}.

Near the magic angle, the quenching of the kinetic energy induced by non dispersive (flat) bands
can strongly enhance the role of the electronic interactions, especially
if the Fermi energy is close to the van Hove singularity. A number of recent theoretical studies have addressed this topic \cite{Letal18,IS18,IYF18,LZCY18,SB18,LN18,gonzalez_prl19,HL19,YIL19,Letal19},
suggesting this can lead o the formation of novel strongly-correlated phases.
The proliferation of these singularities, due to the increasing number of band-crossings in the flat bands as we decrease the twist angle in TBG \cite{YIL19,choi_cm19,zhang_cm19,LKLV_cm19},
can further enhance van Hove peaks in the DOS.
In addition, the strong filling dependence of the electronic compressibility observed in Ref.~\cite{tomarken_cm19}
and the experimental evidence of filling-induced symmetry-broken phases \cite{efetov_cm19}
emphasize the role of the carrier density in driving the electronic interactions. Note, finally, that the peaks due to van Hove singularities in the density of states can be observed directly by scanning tunneling spectroscopy\cite{kerelsky_cm18,choi_cm19,jiang_cm19,Xetal19}

As shown in Ref.~\cite{GuineaWalet18}, 
the long-range Coulomb interactions in TBG away from CN
may be comparable or larger than the bandwidth,
giving rise to an inhomogeneous Hartree potential which substantially
modifies the shape of the electronic bands.
The dominant role of the Coulomb interaction in the
low energy physics of the TBG near the magic angles has also been discussed out in Refs.~\cite{choi_cm19,LKLV_cm19,jiang_cm19}.

In this work we report on a fully self-consistent Hartree calculation of the
electronic bands of TBG near the magic angle, upon varying the filling of the conduction bands away from CN.
In agreement with the findings of the Ref.~\cite{GuineaWalet18},
we find that the shape, width, and position of the bands near the neutrality point depend significantly on the filling of the same bands.
Remarkably, we find an approximate pinning of the DOS at the Fermi energy,
where the self-consistent bands display the tendency to flatten.
As mentioned above, the proximity of the van Hove singularity
to the Fermi level can lead to enhanced electron-electron interactions. The results reported here suggest that a fine tuning of the carrier density is not required for achieving an enhanced density of states at the Fermi level.
Lattice relaxation leads to the reduction in size of the $AA$ regions \cite{koshino_prb17,GuineaWalet_prb19}. It is taken into account approximately in our calculations by assuming the interlayer hopping between $AA$ and $BB$ sublattices is lower than the hopping between $AB$ and $BA$ sublattices \cite{KYKOKF18}.

Our findings are valid in a wide range of values for the
carrier density and the dielectric properties of the substrate.
We intended to study the  broken symmetry phases\cite{XM18} in a  future work. 
In this paper we make a comprehensive study of the effects of long-range electrostatic interactions, which  is a crucial prerequisite to the understanding of more complex phases.

The work is organized as follows:
in Sec.~\ref{model} we provide a technical description of the model used,
introducing an inhomogeneous Hartree potential within the continuum model of TBG.
In Sec.~\ref{numerical_results} we show and comment the numerical results concerning
the self-consistent electronic bands and the corresponding DOS for different fillings
and twist angles.
We also provide an exhaustive analysis of the shape of the Fermi surface and
of the positions of the van Hove singularities in the momenta space. In this section we study situations where the electrostatic potential is strongly screened, in order to show better the continuity between non interacting and interacting results. 
In Sec.~\ref{fock} we consider the role of the exchange potential.
We show that its contribution is sub-leading when compared to the Hartree term
and, more importantly, has little effect on the shape of the bands
and on the position of the van Hove singularity.
In Sec.~\ref{bandwidth} we extend the results to cases with more realistic and smaller screening. We find a substantial enhancement of the bandwith. The total bandwidth is further studied in Sec.~\ref{strains} were we complete the analysis by adding the effect of possible residual strains, see Ref.~\cite{bi_cm19}. 
These combined effects might account for a more precise quantitative description of the spectra of TBG. The appendices complete the information in the main text by analyzing further: i) screening by metallic gates, ii) screening by high-energy bands in the bilayer iii) extension of the calculations to a simplified model with infinitely narrow bands at neutrality, where the dispersion is solely determined by electrostatic effects, and iv) a description of the optical conductivity and the contribution from the van Hove singularities, which could be used to study this problem experimentally.


\section{The model}\label{model}
\begin{figure}
\includegraphics[width=2.5in]{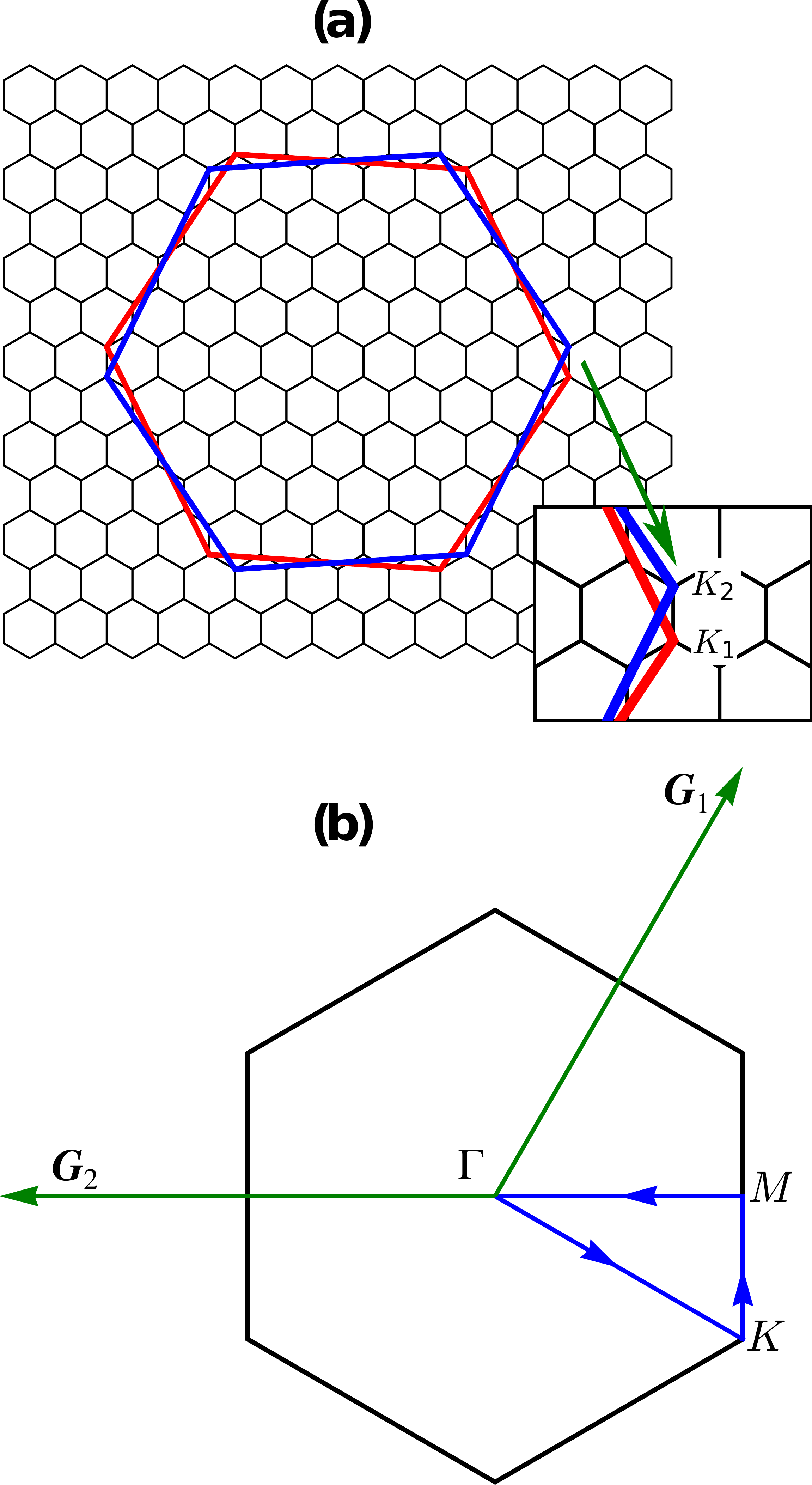}
\caption{
(a) Folding of the BZs of the twisted monolayers graphene.
The BZ of layer 1 (red hexagon) is rotated by $-\theta/2$, while that of the layer 2 (blue hexagon) by $\theta/2$.
The small black hexagons represent the mBZs forming the reciprocal moir\'e lattice.
In the inset: $K_{1,2}$ are the Dirac points of the twisted monolayers, which identify the corners of the mBZ.
(b) mBZ. $\vec{G}_{1,2}$ are the two reciprocal lattice vectors. The blue line shows the high symmetry circuit in the mBZ
used to compute the bands shown in Fig.~\ref{bands}.
}
\label{BZ}
\end{figure}

We model the TBG at small twist angles within the low energy continuum model
considered in  Refs.~\cite{LPN07,BM11,SGG12,LPN12,KYKOKF18}.
At angles  $\theta$ where the two layers are commensurate,
the twist between neighboring layers leads to the formation of a moir\'e superlattice.
The size of the corresponding unit cell \begin{equation}
L=\frac{a}{2\sin\theta/2}\end{equation}
 dramatically increases
as the angle is reduced, where $a=2.46${\AA} is the lattice constant of monolayer graphene.
The continuum models are meaningful for small enough angles, where any
twist forms approximately commensurate structures.
The moir\'e mini Brillouin zone (mBZ), resulting from the folding of the two BZs of each monolayer
(see Fig.~\ref{BZ}(a)),
has the two reciprocal lattice vectors
\begin{equation}
\vec{G}_1=2\pi(1/\sqrt{3},1)/L\text{ and } \vec{G}_2=4\pi(-1/\sqrt{3},0)/L,
\end{equation}
 shown in green in Fig.~\ref{BZ}(b).

For small angles of rotation the coupling between different valleys of the two monolayers
can be safely neglected, as the interlayer hopping has a long wavelength modulation.
In what follows we describe the model for the behavior of the two $K$-valleys of the twisted monolayers,
where $K=4\pi(1,0)/3a$ is the Dirac point of the unrotated monolayer graphene.
The case corresponding to the opposite valleys at $K'=-K$ directly follows from time reversal symmetry
$\vec{k}\rightarrow-\vec{k}$.

The fermionic field operators of the TBG are  4-component Nambu spinors:
\bea
\Psi=\left(\psi_{A_1},\psi_{B_1},\psi_{A_2},\psi_{B_2}\right)^T,
\eea 
where $A,B$ and $1,2$ denote the sub-lattice and layer indices, respectively.
We introduce a relative twist $\theta$ between the two monolayers 
by rotating the layer $1$ by $-\theta/2$ and the layer $2$ by $\theta/2$.
Without loss of generality, we assume that in the aligned configuration, at $\theta=0$,
the two layers are $AA$-stacked.
In the continuum limit, the effective Hamiltonian of the TBG can be generally written as
\cite{LPN07,BM11,KYKOKF18}:
\bea\lb{HTBG}
H_{TBG}=\int\,d^2\vec{r}\Psi^\dagger(\vec{r})
\begin{pmatrix}
H_1&U(\vec{r})\\U^\dagger(\vec{r})&H_2
\end{pmatrix}
\Psi(\vec{r}).
\eea
Here
\bea
H_l=\hbar v_F \vec{\sigma}_{\theta_l/2}\cdot\left(-i\vec{\nabla}-K_l\right)
\eea
is the Dirac Hamiltonian for the layer $l$, where $v_F=\sqrt{3}ta/(2\hbar)$ is the Fermi velocity,
$t$ is the hopping amplitude between $p_z$ orbitals located at nearest neighbors carbon atoms,
$\theta_{1,2}=\mp\theta$,
$\vec{\sigma}_{\pm\theta/2}=e^{i\sigma_z\theta/4}\vec{\sigma}e^{-i\sigma_z\theta/4}$,
$\sigma_i$ are the Pauli matrices, and $K_{1,2}=4\pi\left(\cos\theta/2,\mp\sin\theta/2\right)/(3a)$
are the Dirac points of the two twisted monolayers, which identify the corner of the mBZ shown in
Fig.~\ref{BZ}(a).
The inter layer potential, $U(\vec{r})$, is a periodic function in the moir\'e supercell.
In the limit of small angles, its relevant harmonics are given by the three reciprocal lattice vectors
$\vec{G}=0,-\vec{G}_1,-\vec{G}_1-\vec{G}_2$ \cite{LPN07}.
The corresponding amplitudes $U\left(\vec{G}\right)$ are given by:
\bea
U(0)&=&\begin{pmatrix}g_1&g_2\\g_2&g_1\end{pmatrix},\nn\\
U\left(-\vec{G}_1\right)&=&\begin{pmatrix}g_1&g_2e^{-2i\pi/3}\\g_2e^{2i\pi/3}&g_1\end{pmatrix},\\
U\left(-\vec{G}_1-\vec{G}_2\right)&=&\begin{pmatrix}g_1&g_2e^{2i\pi/3}\\g_2e^{-2i\pi/3}&g_1\end{pmatrix}.\nn
\eea
In the following we adopt the parametrization of the TBG
given in the Ref.~\cite {KYKOKF18}: $\hbar v_F/a=2.1354$ eV,
$g_1=0.0797$ eV and $g_2=0.0975$ eV.
The difference between $g_1$ and $g_2$, as described in Ref.~\cite {KYKOKF18},
accounts for the corrugation effects where the interlayer distance is minimum at the $AB/BA$ spots
and maximum at $AA$ ones, or can be seen as a model of a more complete treatment of lattice relaxation \cite{GuineaWalet_prb19}.
The Hamiltonian Eq.~\pref{HTBG} then hybridizes states of layer $1$ with momentum
$\vec{k}$ close to the Dirac point with the states of layer $2$ with momenta
$\vec{k},\vec{k+G_1},\vec{k+G_1+G_2}$.
For the parametrization used here we find a pair of flat bands, with vanishing Fermi velocity and no dispersion near the Fermi level
for a magic angle $\theta=1.08^\circ$.

Next, we take into account the long-range Coulomb interaction by adding an additional
Hartree term in the Hamiltonian of the Eq.~\pref{HTBG}.
As emphasized in Ref.~\cite{GuineaWalet18} by means of a Wannier-based analysis
(see also \cite{po_prx18,KYKOKF18,kang_prx18}),
the inhomogeneous charge distribution
in TBG leads to an electrostatic potential of the order of $V_{\text{Coulomb}}\sim e^2/\epsilon L$, where $e$ is the electron charge and $\epsilon$ the dielectric constant of the substrate. At angles $\theta\sim1^\circ$, $V_{\text{Coulomb}}$
turns out to be generally much bigger than both the mini-band width, which is a few meV, 
and the intra-atomic Hubbard interaction,
and consequently strongly modifies the band structure close to the Fermi level.
In general the Hartree contribution to the Hamiltonian is given by
\bea
V_H&=&\int\,d^2\vec{r}v_H(\vec{r})\Psi^\dagger(\vec{r})\Psi(\vec{r}),\\
v_H(\vec{r})&=&\int\,d^2\vec{r}'v_C(\vec{r}-\vec{r}')\d \rho(\vec{r}').\lb{vH}
\eea
Here $v_C(\vec{r})=\frac{e^2}{\epsilon \left|\vec{r}\right|}$ is the Coulomb potential and $\d \rho(\vec{r})$
is the fluctuation of the charge density with respect to CN.
It is worth noting that, in the continuum model, the charge density does depend on the cutoff and,
consequently, a convention has to be applied. In what follows we impose the condition that the Hartree potential vanishes at CN, where $\delta\rho=0$ by definition.
The density of carriers in TBG is periodic in the moir\'e unit cell,
with maxima in the $AA$ spots and minima in the $AB/BA$ ones
\cite{SGG12,LPN12,GuineaWalet18,TKV_cm19}.
Thus  it can be generally written as the superposition of plane waves:
\bea
\delta\rho(\vec{r})=\frac{1}{V_c}\sum_{\vec{G}}\delta\rho(\vec{G})e^{i\vec{G}\cdot\vec{r}},
\eea 
where $V_c=L^2\sqrt{3}/2$ is the area of the unit cell and the $\vec{G}$'s are reciprocal lattice vectors.
The amplitudes $\delta\rho(\vec{G})$ are given by:
\bea\lb{drhoG}
\delta\rho(\vec{G})=4\int_{\text{mBZ}}\frac{\,d^2\vec{k}}{V_\text{mBZ}}\sum_{\vec{G'}i,l} 
\phi^{i,*}_{l\vec{k}}\left(\vec{G'}\right)\phi^{i}_{l\vec{k}}\left(\vec{G'}+\vec{G}\right),
\eea
where $V_\text{mBZ}=(2\pi)^2/V_c$ is the area of the mBZ,
$l$ is the band index resulting from the Bloch diagonalization of the full Hamiltonian,
$H_{TBG}+V_H$. The parameter $i=A_1,B_1,A_2,B_2$ labels the sub lattice and layer
and $\phi^i_{l\vec{k}}\left(\vec{G}\right)$
is the amplitude for an electron to occupy a state with Bloch momentum $\vec{k+G}$
in the $l$-th band. The $\phi$'s are normalized  as
\bea\sum_{i\vec{G}}\phi^{i,*}_{l\vec{k}}\left(\vec{G}\right)\phi^i_{l'\vec{k}}\left(\vec{G}\right)=\delta_{l,l'}.\eea
In Eq.~\pref{drhoG}, the sum over the band index $l$ extends only over  energy levels with  $E_{l}(\vec{k})$ between
the CN point and the Fermi level. The factor of $4$
accounts for the  spin/valley degeneracy of the TBG Hamiltonian.
The Hartree potential of the Eq.~\eqref{vH} implicitly depends on the fractional filling of the conduction band:
$n\in[-4,4]$, where $n=4$ corresponds to fully filled valence and conduction bands,
while they are both empty for $n=-4$.
 
As shown in  Ref.~\cite{GuineaWalet18},
the dominant contribution to $\d \rho$
comes from the first star of reciprocal lattice vectors, 
$\vec{G}=\vec{G}_i=\pm\vec{G}_1,\pm\vec{G}_2,\pm\left(\vec{G}_1+\vec{G}_2\right)$.
As a consequence, we will neglect higher order harmonics in the following.
In addition, exploiting  $C_6$ symmetry
we can assume that the Fourier components of $\d\rho$
are equally weighted in the first star of vectors $\vec{G}_i$.
Under these assumptions, the Hartree potential of the Eq.~\pref{vH} takes the form:
\bea\label{vH2}
v_H(\mathbf{r})=
V_0\d \rho_G\sum_i e^{i\vec{G}_i\cdot\vec{r}},
\eea 
where $V_0=v_C\left(\vec{G}_i\right)/V_c$, with
$v_C\left(\vec{G}_i\right)=2\pi e^2/\left(\epsilon\left|\vec{G}_i\right|\right)$
 the Fourier transform of $v_C$ evaluated at $\vec{G}_i$,
and thus $V_0=e^2/\left(\epsilon L\right)$.
The real quantity
\bea\label{rhoG}
\d\rho_G\equiv \frac{1}{6}\sum_i\int_{V_c}\,d^2\vec{r}\delta\rho(\vec{r})e^{-i\vec{G}_i\cdot\vec{r}}
\eea
is the single parameter which defines the Hartree potential, and thus needs to be determined self-consistently.
Note that the Hartree potential \eqref{vH2} contributes to the Hamiltonian
of the TBG as a diagonal term in the sub-lattice/layer subspace,
coupling the generic state of momentum $\vec{k}$ to the six states of momentum
$\vec{k}-\vec{G}_i$,
with an amplitude equal to $V_0\delta\rho_G$.

For a given value of the filling of the conduction band, $n$,
we compute the valence and conduction bands
by diagonalizing the Hamiltonian for the TBG
in the presence of the Hartree potential of Eq.~\pref{vH2}, $v_H$,
in which $\d\rho_G$ is obtained as the self-consistent solution of the Eq.~\pref{rhoG}.
This procedure is iterated until convergence.
Numerical details concerning the iterative method are provided in Appendix \ref{appA}.

\section{Bands distortion induced by the inhomogeneous Hartree potential}\label{numerical_results}
In this section we show the numerical results for the self-consistent diagonalization described in the previous section,
as a function of the twist angle and the filling $n$.

The self-consistent order parameter, $\delta\rho_G$,
is a dimensionless number varying between $-4$ and $4$.
As shown in Ref. \cite{GuineaWalet18} and in Fig. \ref{rhoG_vs_n} below,
the value of $\delta\rho_G$ depends approximately linearly on $n$
in a wide range of fillings and parameters.
For the results in this section, we use a very  high value of the dielectric constant $\epsilon\simeq 66$. to reduce the strength of the Coulomb force.
For this choice  we find that the amplitude of the fluctuations of the Hartree potential between the
$AA$ and $AB/BA$ spots of the moir\'e unit cell, 
$V_H(AA)-V_H(AB) \simeq 5\,\mathrm{meV}$, for a band filling $n=1$.
This energy scale is  comparable to the bandwidth of the TBG in the absence of interactions, so that a simple one-to-one comparison between the two cases is possible.
\begin{figure}
\includegraphics[width=.6\columnwidth]{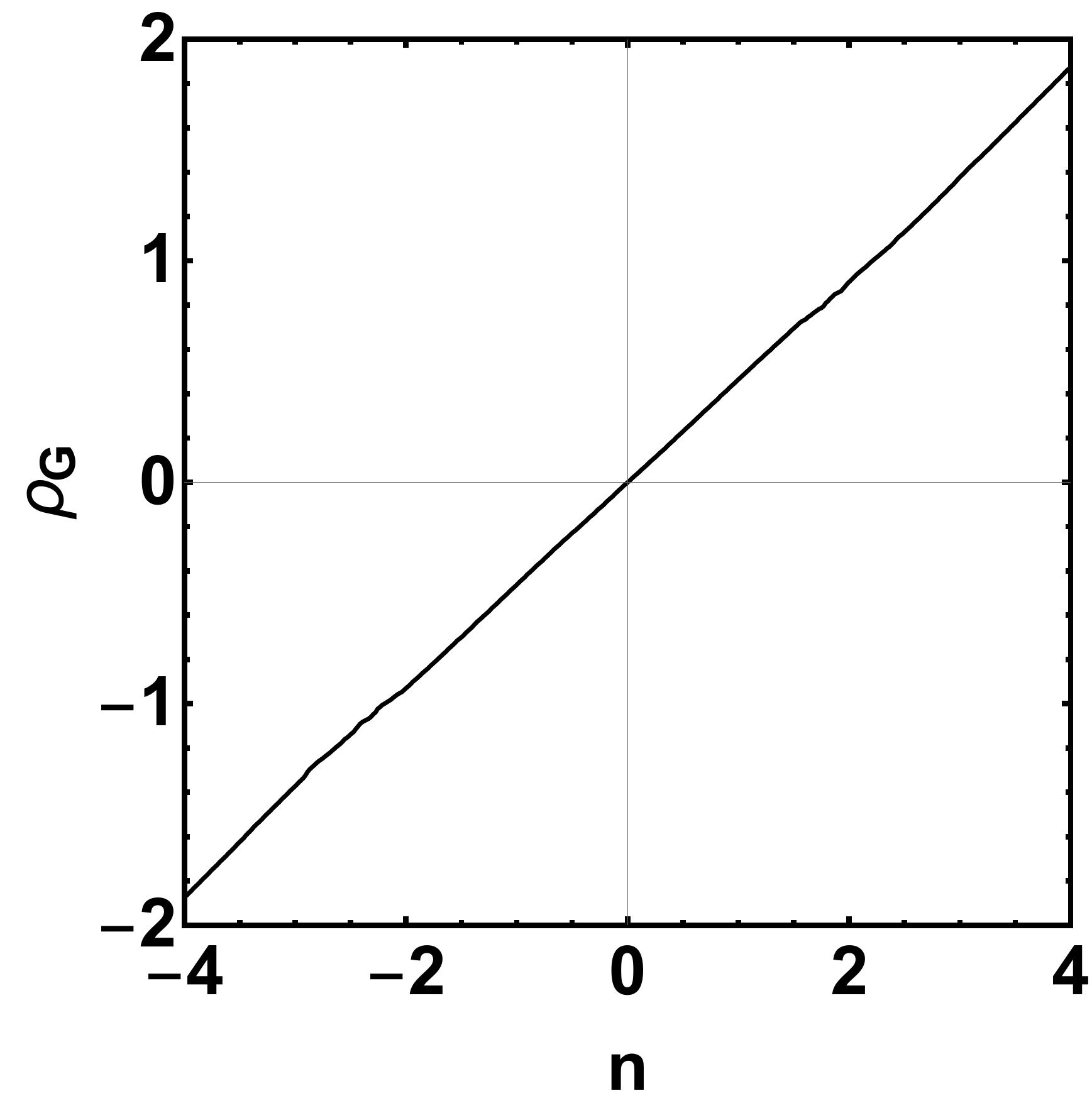}
\caption{
Values of $\delta\rho_G$ as a function of the filling, $n$, obtained for $\theta=1.05^\circ$.
}
\label{rhoG_vs_n}
\end{figure}

A study of the effects of more realistic electrostatic interactions that include a more sensible  screening parameter,  $\epsilon\sim 4-8$, is carried out in Sec.~\ref{bandwidth}. The details  of the screening due to external metallic gates, and to the high-energy subbands in the bilayer is discussed in Appendix A.

Figure~\ref{bands} shows the bands obtained along the high-symmetry circuit of the mBZ, as shown in  Fig.~\ref{BZ}(b),
for various fillings and angles near the magical one.
As we anticipated, the bands do not rigidly shift when varying the filling,
but they rearrange their shape in a quite non trivial way.
Most interestingly they tend to flatten in the vicinity of the Fermi level.
We note that the bands shown in Fig. \ref{bands}b at CN (red line) behave rather differently than those in the other panels.
This is because $\theta=1.08^\circ$ is a magic angle for our choice of parameters,
so that the corresponding low energy bands minimize the bandwidth as compared to the other angles considered in Fig. \ref{bands}.
Also the appearance of new Dirac points is a typical feature of the bands of TBG for angles near the magic one. 
\begin{figure}
\includegraphics[width=\columnwidth]{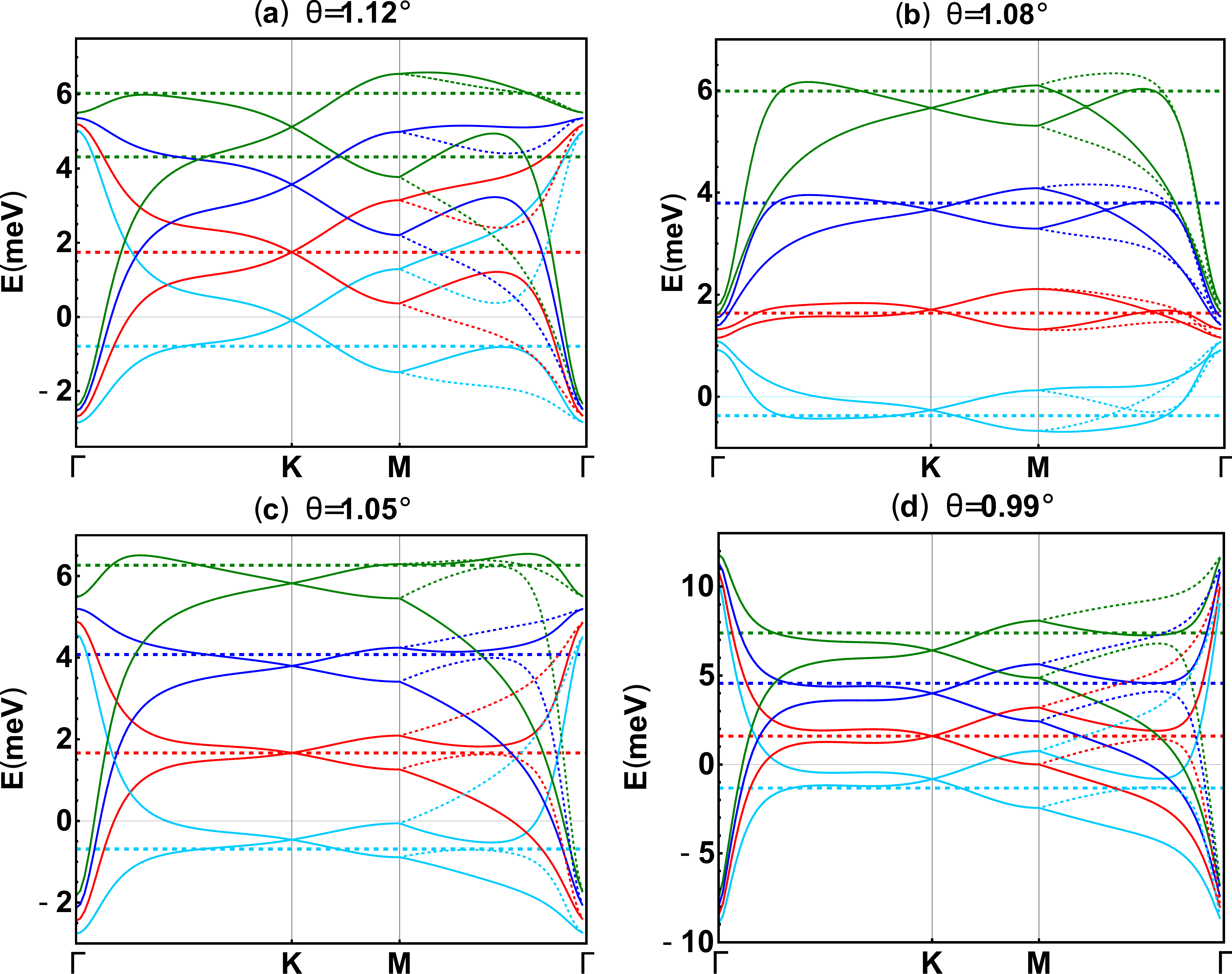}
\caption{
Self-consistent bands of TBG in the presence of the Hartree potential,
obtained for the twist angles $\theta=1.12^\circ,1.08^\circ,1.05^\circ,0.99^\circ$,
and fillings $n=-1$ (cyan), $n=0$ (red), $n=1$ (blue), $n=2$ (green). 
The dotted lines show the bands corresponding to the opposite valleys,
while the horizontal dashed lines represent the Fermi energies.
}
\label{bands}
\end{figure}
This is confirmed by the DOS, shown in Fig.~\ref{dos},
which is clearly pinned at the Fermi energy for $n\ne0$.
\begin{figure}
\includegraphics[width=\columnwidth]{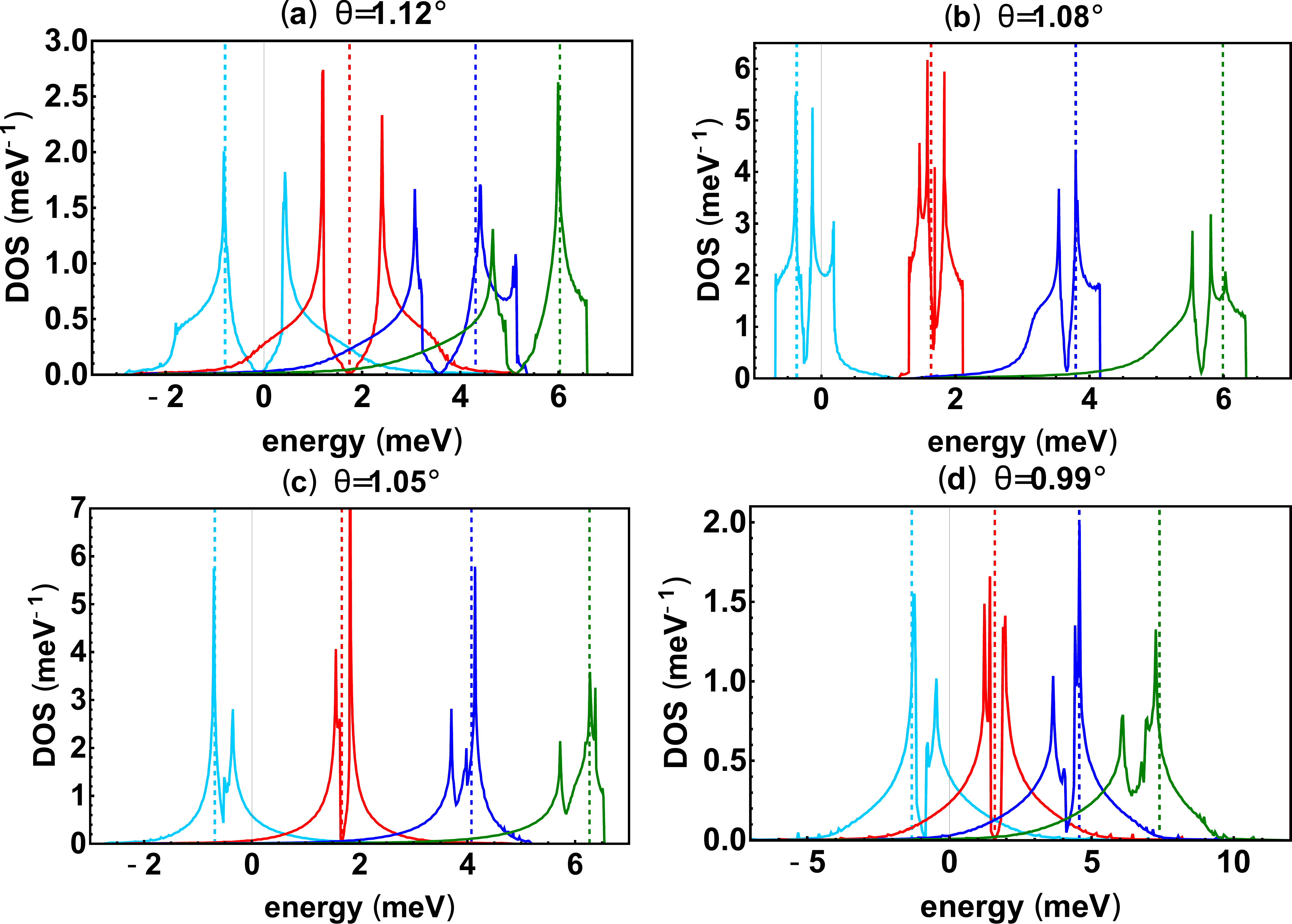}
\caption{
DOS for each of the results shown Fig.~\ref{bands}, color-coded as in that figure.
The vertical dashed lines represent the Fermi energy for each case.
}
\label{dos}
\end{figure}
Moreover, the reshaping of the bands is much more evident around the points $K$ and $M$
than at $\Gamma$, where the bands are not sensitive to a change in the density of carriers.

\begin{figure}
\includegraphics[width=\columnwidth]{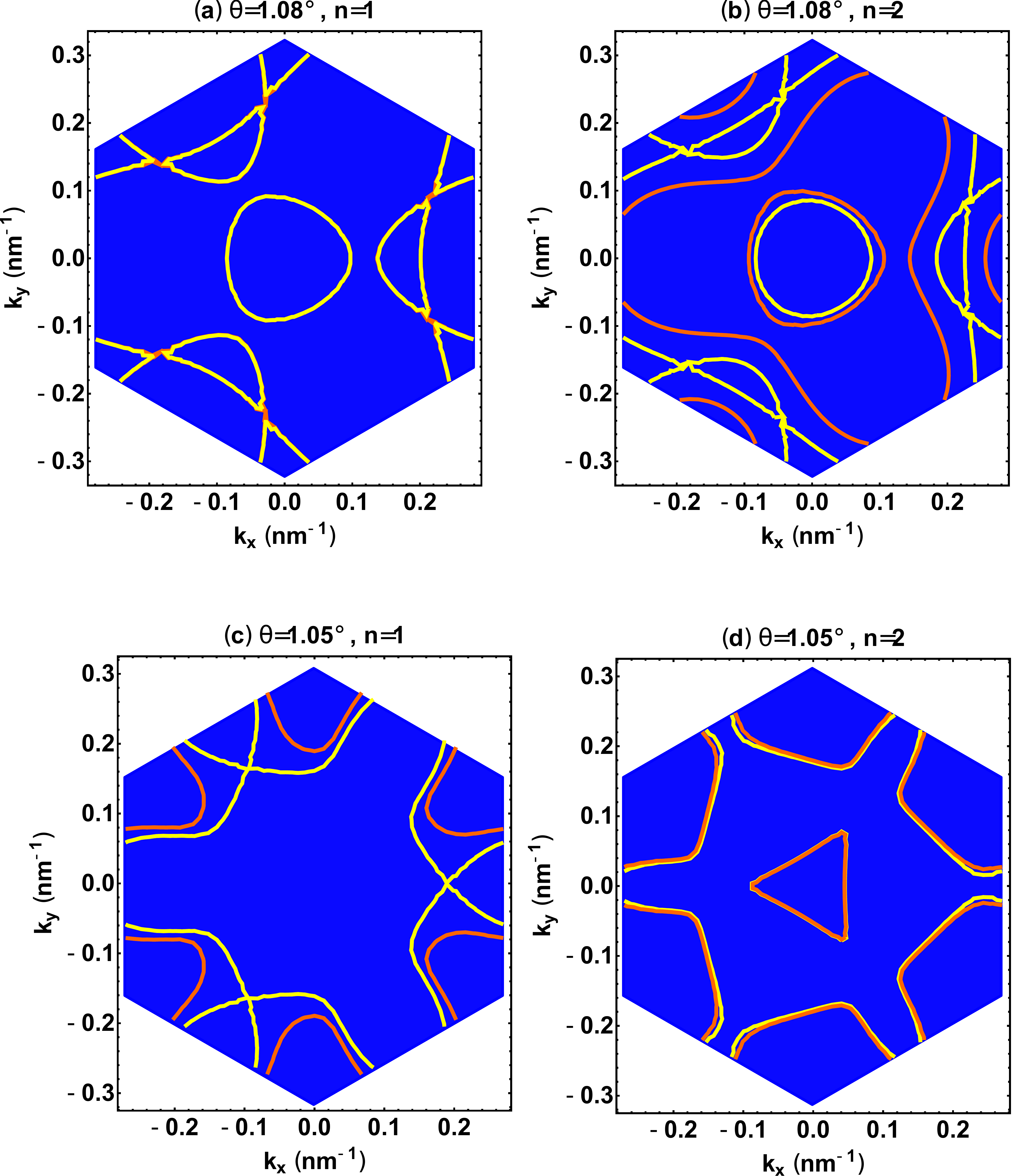}
\caption{
Fermi surfaces (orange)
and positions of the van Hove singularities (yellow)
in the mBZ, for twist angles $\th=1.08^\circ$ and $1.05^\circ$ and filling fraction $n=1,2$.
}
\label{FS}
\end{figure}

In  Fig.~\ref{FS} we can see the details of the  relation of the position of the Fermi surface
and  the van Hove singularities
in the mBZ, for the angles $\th=1.08^\circ,1.05^\circ$ and fillings $n=1,2$.
As already suggested by the results of Figs.~\ref{bands},\ref{dos}, 
the van Hove singularities generally match the Fermi surface closely for many parameters.
This effect is more evident 
for $\th=1.08^\circ$ and $n=1$,
and for $\th=1.05^\circ$ and $n=2$, than for the other two cases.

\begin{figure}
\includegraphics[width=0.8\columnwidth]{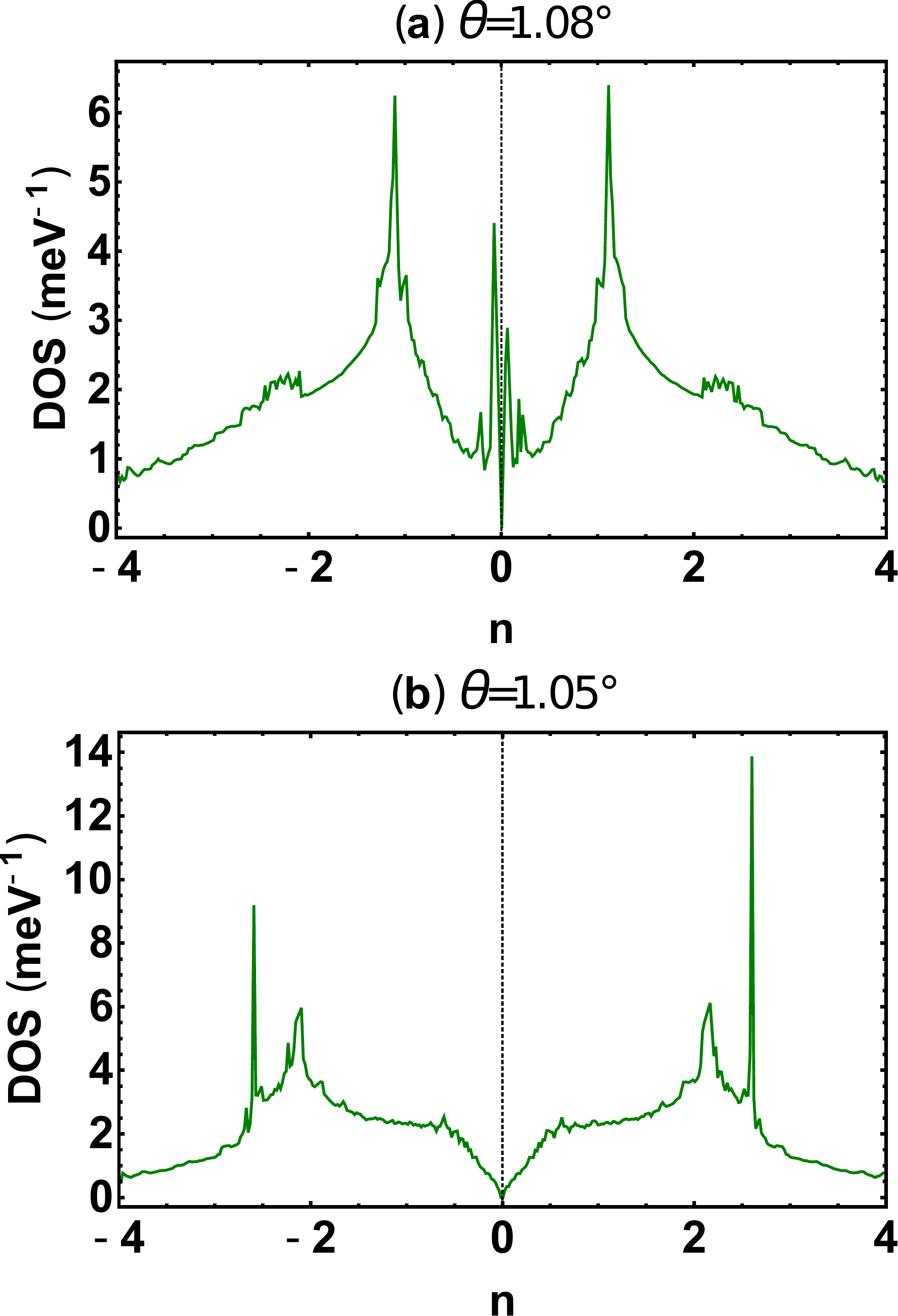} 
\caption{
Value of the DOS at the Fermi level
as a function of the filling of the conduction band,
for the angles $\th=1.08^\circ$ (a) and $\th=1.05^\circ$ (b).
}
\label{DOS_fermi}
\end{figure}

In Fig.~\ref{DOS_fermi} we show the value of the DOS at the Fermi level
as a function of the filling of the conduction band,
for the angles $\th=1.08^\circ$ and $\th=1.05^\circ$.
In both  cases the DOS displays intense isolated peaks.
This  feature
suggests the possibility of enhancing or suppressing the electron-electron interactions in TBG
by fine tuning the number of charge carriers, as recently pointed out in experiments
\cite{tomarken_cm19,efetov_cm19,choi_cm19,jiang_cm19}.
Furthermore, the features of the DOS are generally not particle-hole symmetric,
which agrees with the experimental findings of  Ref.~\cite{tomarken_cm19}.

\begin{figure}
\includegraphics[width=1.\columnwidth]{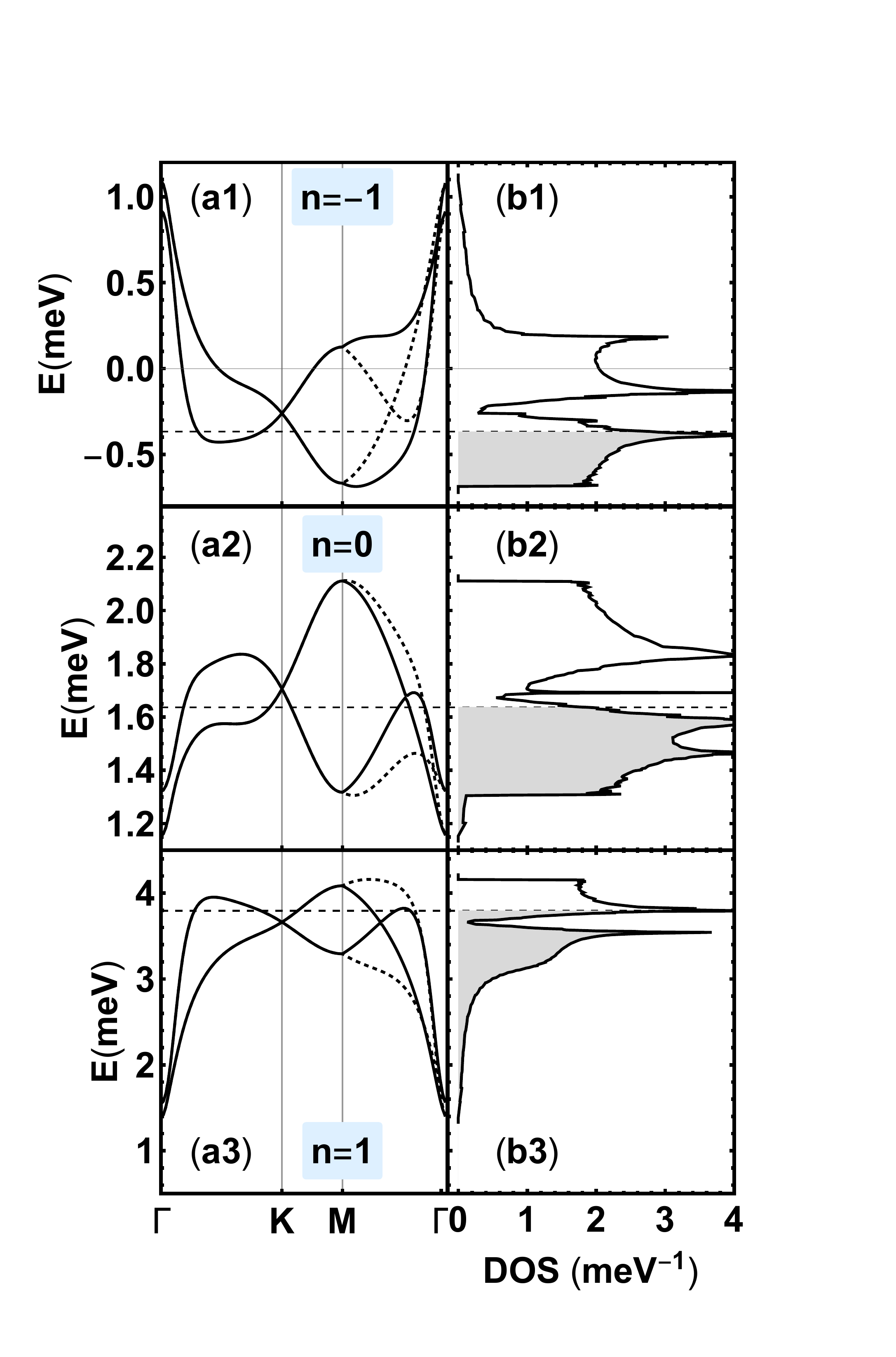} 
\caption{
band structure (panels (a1)-(a3)) and DOS (panels (b1)-(b3)) obtained at the magic angle $\theta=1.08^\circ$ and filling $n=-1,0,1$, as indicated in the legends. The gray area in the plots of the DOS represents the occupied fraction of the spectral weight.
}
\label{1p08deg_filling}
\end{figure}
To gain a more detailed understanding of the evolution of the band structure with filling, and to stress the main differences between the interacting and non-interacting case, we show the bands (Fig. \ref{1p08deg_filling}(a)) and the DOS (Fig. \ref{1p08deg_filling}(b)) obtained at the magic angle $\theta=1.08^\circ$, for the fillings $n=-1,0,1$, where $n=1$ represents the non-interacting case.

\section{Role of the exchange potential}\label{fock}
If we wish to perform a fully consistent mean-field calculation, we need to include the Fock term.
The Fock or exchange potential can be written as
\bea\lb{H_F}
V_F&=&-\sum_{ij}\int\,d^2\vec{r}\,d^2\vec{r'}
v_c\left(\vec{r-r'}\right)\\
&&\times
\left\langle\Psi^{i,\dagger}(\vec{r})\Psi^j(\vec{r}')\right\rangle
	\Psi^{j,\dagger}(\vec{r'})\Psi^i(\vec{r})	 ,\nn
\eea
where the brackets, $\left\langle\dots\right\rangle$, denote the expectation value in the filled states of the full Hamiltonian
$H_{TBG}+V_H+V_F$.
Using an expansion in Bloch waves, we can write this as
\begin{widetext}
\bea\lb{correlator}
\left\langle\Psi^{i,\dagger}(\vec{r})\Psi^j(\vec{r}')\right\rangle=
\int_{\text{mBZ}}\frac{\,d^2\vec{k}}{\left(2\pi\right)^2}
\sum_{\vec{G,G'}}e^{i\vec{k}\cdot(\vec{r'-r})}
e^{i\vec{G'}\cdot\vec{r'}}e^{-i\vec{G}\cdot\vec{r}}
C_{\vec{k}}^{ji}\left(\vec{G'},\vec{G}\right).
\eea
\end{widetext}
Here
\bea
C_{\vec{k}}^{ji}\left(\vec{G'},\vec{G}\right)\equiv
\sum_{l_{occ}}\phi^j_{l\vec{k}}(\vec{G'})\phi^{i,*}_{l\vec{k}}(\vec{G}),
\eea
with the sum  over all the occupied levels,
$l_{occ}$, in the two bands in the middle of the spectrum.
The Hamiltonian \pref{H_F} is more conveniently expressed in momentum space.
Substituting  Eq.~\pref{correlator} into \pref{H_F} and folding the BZ
of monolayer graphene gives
\begin{widetext}
\bea\lb{V_F2}
V_F=-\int_{\text{mBZ}}\,d^2\vec{k}\sum_{ij}\sum_{\vec{G,G'}}
\Psi^{j,\dagger}_{\vec{k}}(\vec{G'})\Psi^i_{\vec{k}}(\vec{G'+G})
v_{F,\vec{k}}^{ji}\left( \vec{G'},\vec{G'+G} \right).
\eea
\end{widetext}
Here
$
\Psi_{\vec{k}}(\vec{G})\equiv \int\frac{\,d^2\vec{r}}{2\pi}\Psi(\vec{r})e^{i\left(\vec{k+G}\right)\cdot{r}}
$
is the Fourier component of the field operator $\Psi(\vec{r})$, corresponding to the Bloch momentum $\vec{k}+\vec{G}$,
and
\begin{widetext}
\bea\lb{VF_ij}
v_{F,\vec{k}}^{ji}\left( \vec{G'},\vec{G'+G} \right)\equiv
-\sum_{\vec{G''}}\int_{\text{mBZ}}\frac{\,d^2\vec{p}}{V_{\text{mBZ}}}
\frac{v_c\left(\vec{p-k+G''}\right)}{V_c}
C^{ji}_{\vec{p}}\left(\vec{G''+G',G''+G'+G}\right).
\eea
\end{widetext}
The off-diagonal elements, $v_{F}^{j\ne i}$, can be safely neglected,
as the interactions coupling different layers and/or sub lattices
are supposed to be local in space within the continuum model,
while the exchange potential describes the non-local effects of the Coulomb interaction.
Assuming that the Coulomb potential is negligible at momenta outside the mBZ,
we retain only the term corresponding to $\vec{G''}=0$ in the sum of the Eq.~\pref{VF_ij}.
In addition, to link with the truncation of the Hartree potential,  we only consider  matrix elements of the exchange potential
\pref{V_F2}
coupling states of momentum $\vec{k+G}$ with states of momentum $\vec{k+G}$
and $\vec{k+G+G_i}$, where
$\vec{G_i}$ belongs to the first star of reciprocal lattice vectors.

\begin{figure}
\includegraphics[width=3.5in]{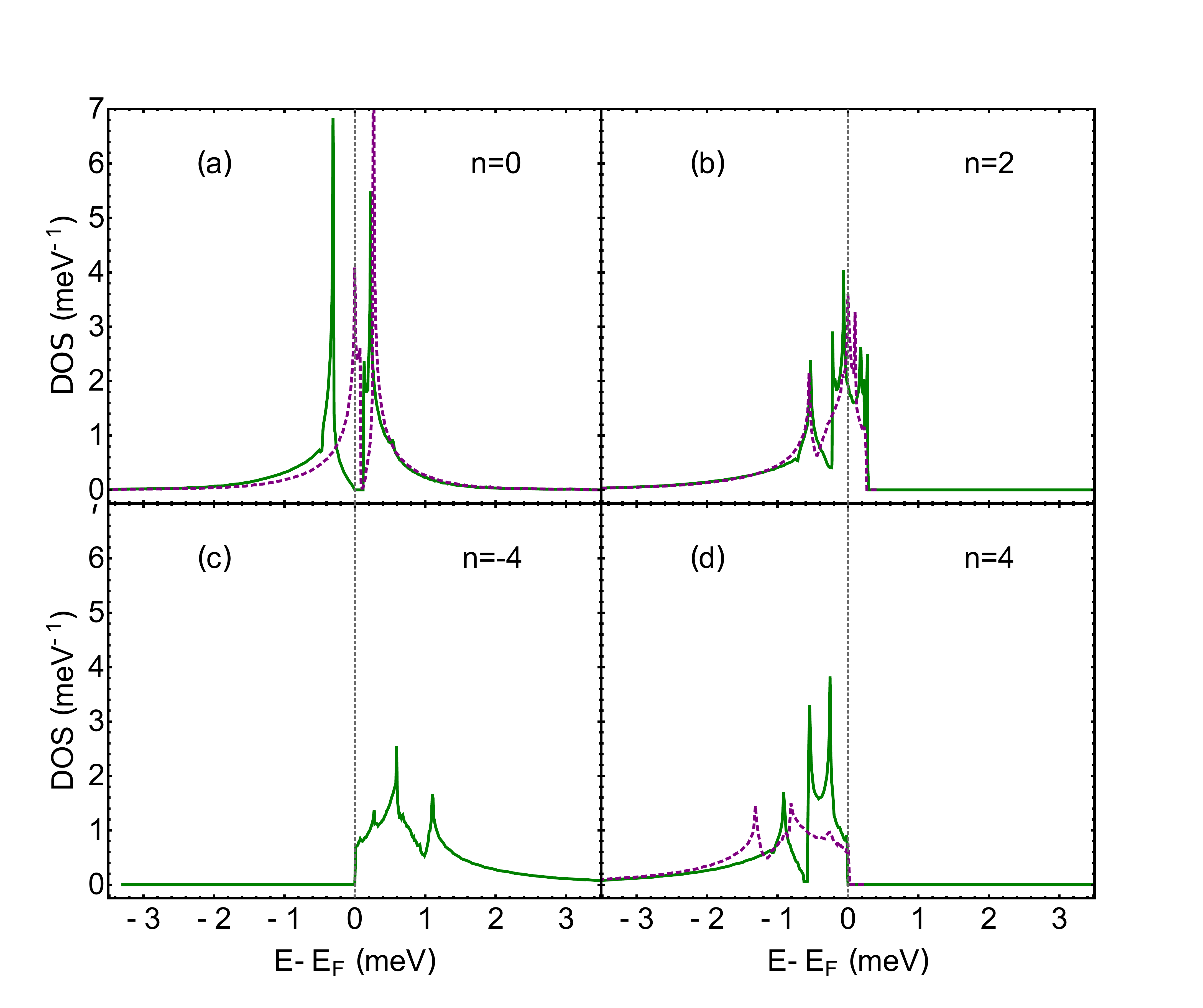}
\caption{
Density of states obtained for a twist angle $\theta=1.05^\circ$ including the full Hartree-Fock interaction (solid lines)
and only the Hartree part (dotted lines). The origin of the horizontal axis has been set at the Fermi level.
}
\label{dos_HF}
\end{figure}

In Fig.~\ref{dos_HF} we compare the DOS obtained using the Hartree-Fock approximation  (continuum lines)
to those with only Hartree potential (dotted lines), for a twist angle $\theta=1.05^\circ$ and
different fillings of the conduction band. 
The main effect of the exchange potential is visible  at CN, $n=0$, where the Hartree potential
is absent, and  at full filling, $n=4$, where the exchange potential is enhanced by the contribution of all the electronic states.
However, in both these cases the difference between the results corresponding to the full Hartree-Fock calculations and those obtained with only the Hartree potential
is not qualitatively significant, even if the DOS differs quantitatively in the two cases.
In panel (b), for two electrons per unit cell in the conduction bands (including valley and spin), the Fock contribution is negligible when compared to
the Hartree one. For  empty bands,  panel (c), the exchange potential is zero.
We can conclude that adding the exchange potential only makes minor modifications to the DOS or the
pinning at the Fermi level, but does not give a qualitative change to the main results of this work.
We conclude that the main effect causing distortion of the bands originates from the Hartree potential.   

\section{Effect of screening}\label{bandwidth}

As we mentioned above, the results shown in the previous sections
correspond to a strong screening, in order to facilitate the comparison between the non-interacting and the interacting bands. The screening from metallic gates, and from the higher energy bands of the bilayer, at the relevant wavevector for the electrostatic potential, $| \vec{ G} | = ( 4 \pi )/( \sqrt{3} L_M )$, is not expected to be very large. Typical values for the dielectric function of the hBN slabs that encapsulate the bilayer are $\epsilon \approx 4 - 8$ (note that, in addition, a very thin hBN slab does not screen electrostatic potentials at low wavevectors, see Appendix \ref{appB}. In this section, we analyze the role of the Hartree potential for screening constants of $\epsilon = 4$ and $\epsilon = 7.5$.

\begin{figure}
\includegraphics[width=3.5in]{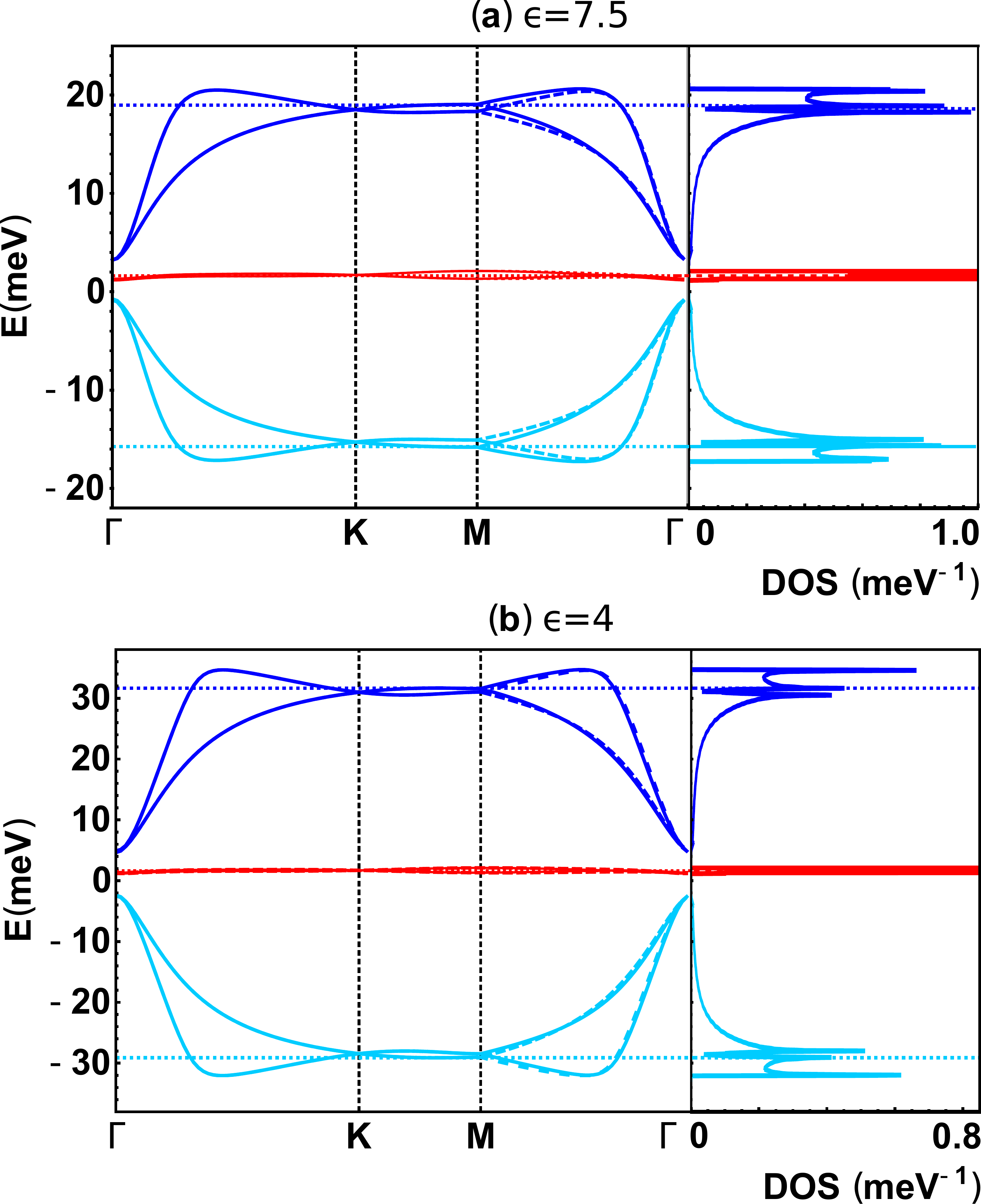}
\caption{
Bands and DOS for $V_0=15\,\mathrm{meV}$ (a) and $V_0=28\,\mathrm{meV}$ (b), at the angle $\th=1.08^\circ$,
and filling $n=-1$ (cyan), $n=0$ (red), $n=1$ (blue).
}
\label{eps_1p08}
\end{figure}

\begin{figure}
\includegraphics[width=3.5in]{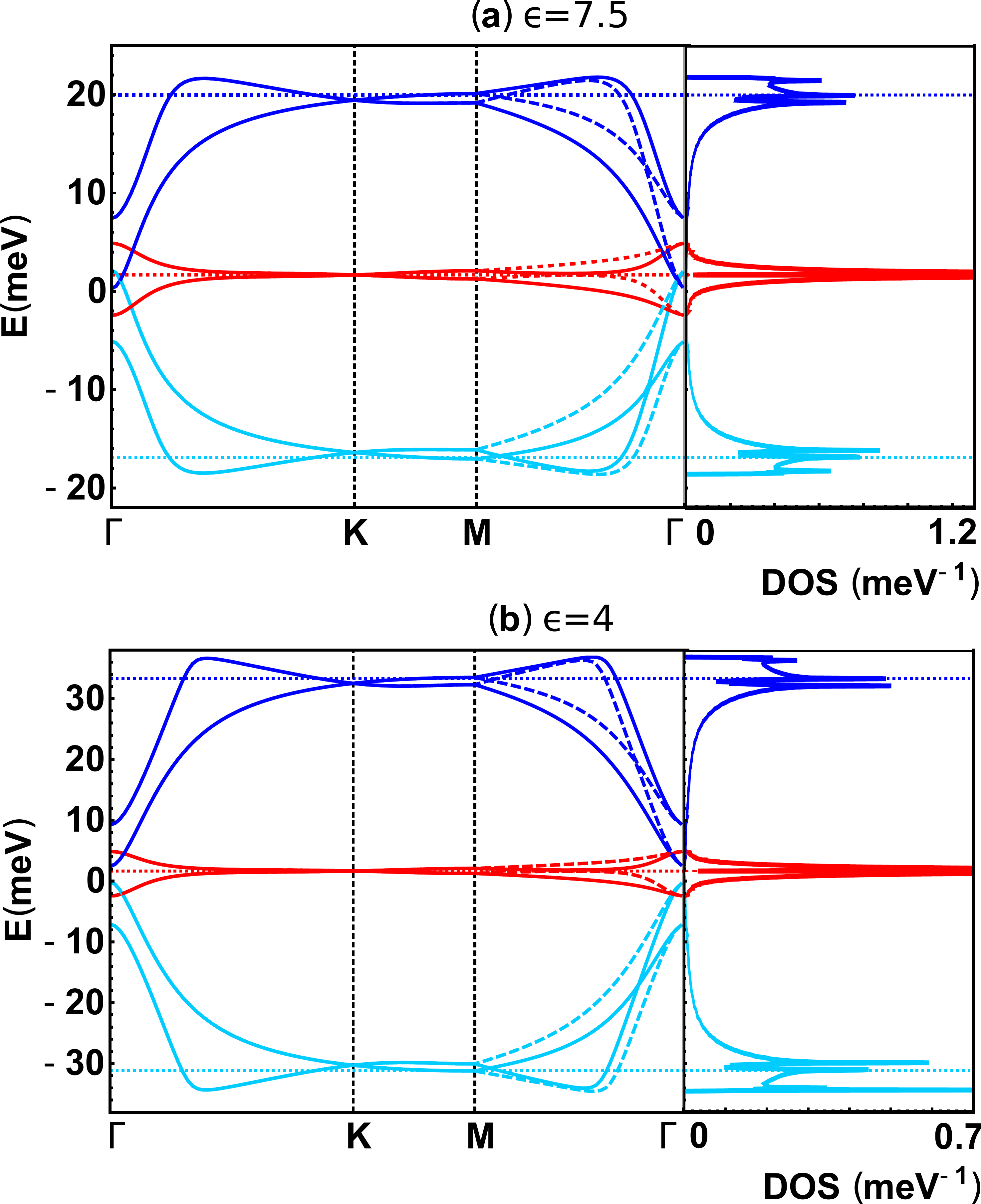}
\caption{
As in Fig.~\ref{eps_1p08}, but for $\th=1.05^\circ$.
}
\label{eps_1p05}
\end{figure}

Fig.~\ref{eps_1p08} shows the bands (left panels) and the DOS (right panels) obtained for $\epsilon=7.5$ (a) and $\epsilon=4$ (b),
for $n=-1,0,1$ and twist angle $\theta=1.08^\circ$. Analogous results corresponding to $\theta=1.05^\circ$ are shown in Fig.~\ref{eps_1p05}.
The distortion of the bands induced by the Hartree potential is much stronger than for the results shown in Sec.~\ref{numerical_results}.
Though, the data still display a clear pinning of the DOS at the Fermi level and the main qualitative features pointed out in the previous sections remain nearly unchanged.
In addition, the overall bandwidth sensitively grows as $\epsilon$ is lowered.
For the case corresponding to $\epsilon=4$, which is a quite realistic value, we obtain a bandwidth of approximately $30\,\mathrm{meV}$.
This value is much larger than what predicted by the continuum models \cite{LPN07,BM11,SGG12,LPN12,KYKOKF18}
($\lesssim5\,\mathrm{meV}$).
Consequently, we argue that the Coulomb interaction will make an important  contribution in broadening the mini-band width of the TBG
near the magic angle.

\section{Effect of residual strains}\label{strains}

\begin{figure}
\includegraphics[width=3.5in]{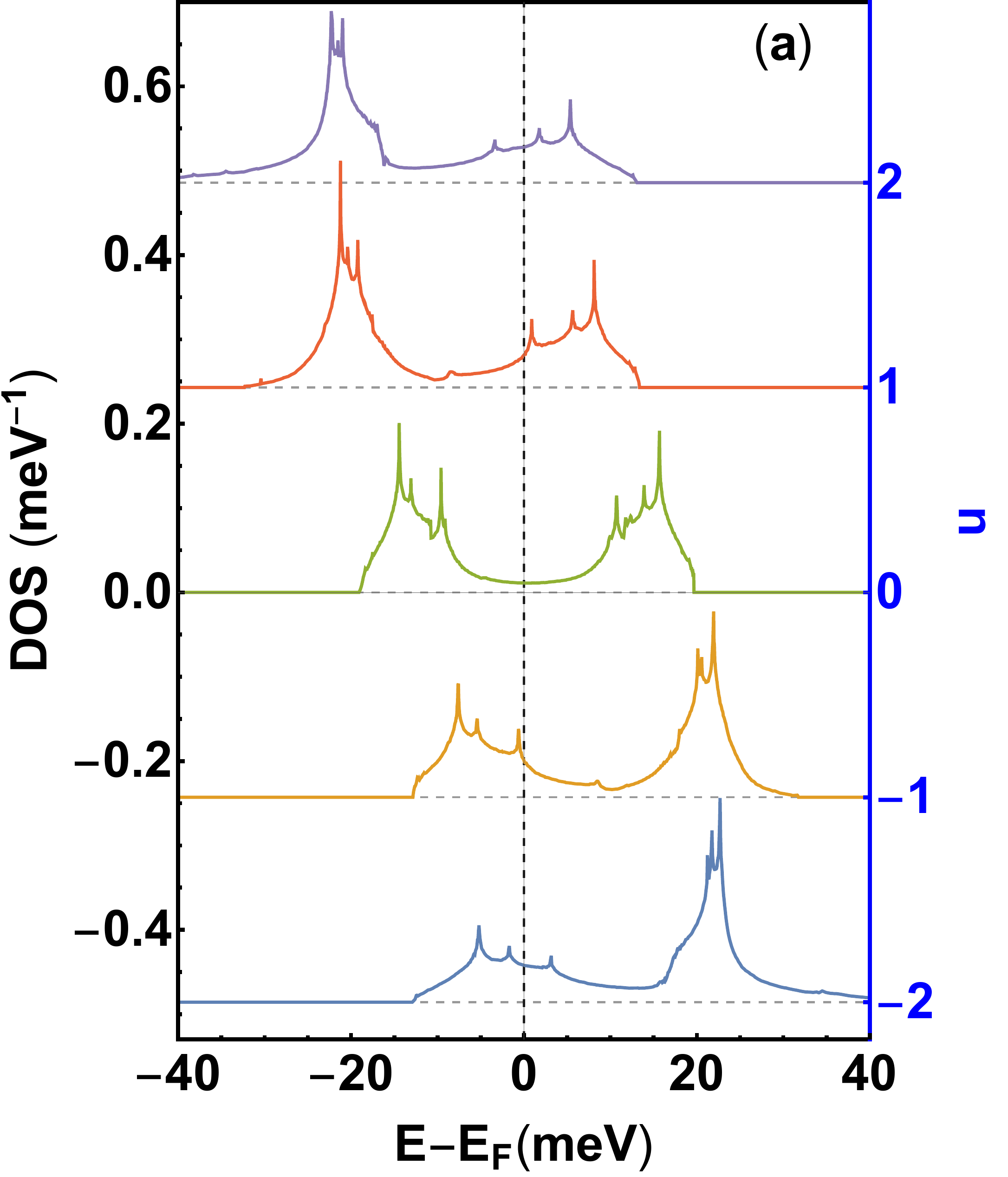}
\includegraphics[width=2.5in]{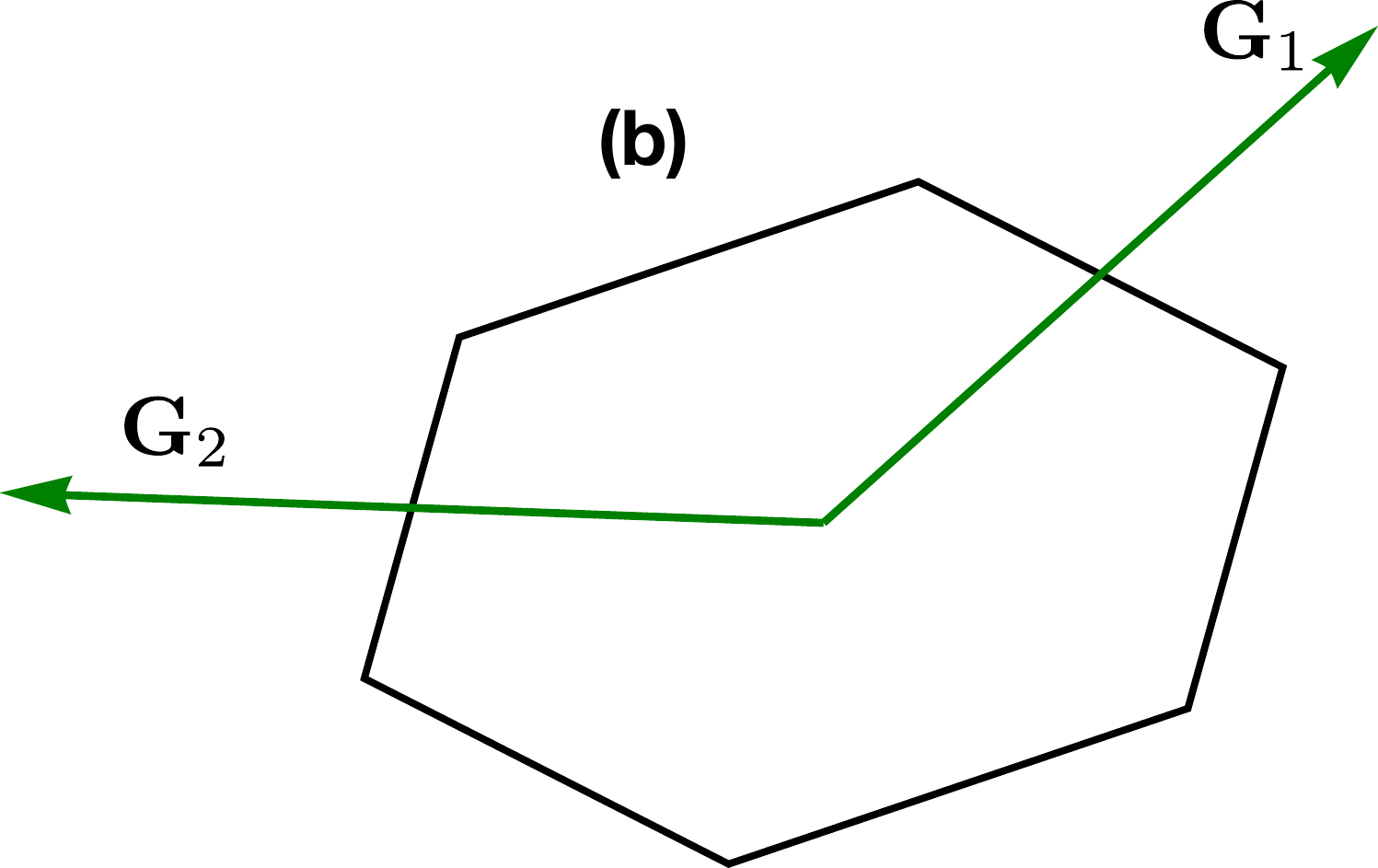}
\caption{
(a) DOS obtained for a twist angle $\theta=1.05^\circ$ including uniaxial strain $\varepsilon=0.6\%$ for a dielectric constant $\epsilon=4$.
The zeroes for each curve (denoted by the dashed lines) have been moved for each filling as indicated on the right-hand side . (b) Distorted mBZ in the presence of strain.
}
\label{strain}
\end{figure}

For completeness, we consider the additional effects of residual strains, which may play a role in realistic situations. The presence of strain is revealed by the elliptical shape of the moir\'e dots observed in STM experiments \cite{kerelsky_cm18}.
As shown in \cite{koshino_prb17,bi_cm19}, the combined effects of relaxation and strain might
explain the large distance between the van Hove singularities, up to $55\,\mathrm{meV}$,
observed in \cite{kerelsky_cm18,tomarken_cm19}.
Here we consider a uniaxial strain of $\varepsilon=0.6\%$, applied along the direction forming an angle of $30^\circ$ with respect to the horizontal axis. This leads to a distortion of the mBZ as shown in Fig.~\ref{strain}(b). For the technical details concerning the parametrization of the strain in the continuum model of TBG, we refer to Ref.~\cite{bi_cm19}. Figure~\ref{strain}(a) shows the DOS obtained including strain and a dielectric constant $\epsilon=4$, for the twist angle $\theta=1.05^\circ$ and different fillings of the conduction bands, as labeled by the vertical right axis.
The overall bandwidth is delimited by the two peaks associated to the van Hove singularities, with a distance between them of approximatively $\sim35\,\mathrm{meV}$, in agreement with the experimental observations \cite{kerelsky_cm18,tomarken_cm19}.
In general, we find a less pronounced pinning of the DOS at the Fermi energy as compared to the strain-less case of the Sec.~\ref{numerical_results}. Interestingly, moving away from CN, the peak closest to the Fermi energy broadens while the farthest one gains intensity.

\section{Conclusions}

We have analyzed the effect of the electrostatic interaction on the band structure of a TBG near the magic angle $\theta\sim1^\circ$, with emphasis on the width of the bands, and the change in the role of the van Hove singularities in the density of states. The scale of the electrostatic potential, $\sim e^2 / \epsilon L_M )$, is comparable to or larger than  the bandwidth $W$ of the non-interacting system. Even for an unrealistically large screening, $\epsilon \sim 50 - 80$, the scale of the Coulomb force equals the band splitting near the first magic angle, suggesting that such effect may drive the electronic properties of such systems. 

An analysis of the screening by the hBN substrate, metallic gates, and high energy bands of the twisted bilayer allows us to estimate that realistic values for the screening of the electrostatic potential are in the range $\epsilon \sim 6 - 10$. For these values of the screening, the bandwidth is mostly determined by the Hartree potential--the exchange terms have little effect, apart for CN where the Hartree terms vanish.
The strong sensitivity of the band structure to the screening by the environment opens new ways of manipulating the system.

The effect of the Hartree potential varies significantly as function of filling.
This dependence of the shape of the bands on electronic filling shifts states from above (below) the van Hove singularities to below (above) them, depending on whether the carriers are electrons or holes. This leads to an approximate pinning of the van Hove singularities at the Fermi energy at certain fillings, in agreement with experimental results.

This suggest that interactions stabilize such pinning, making it more stable over a wider range of fillings, and thus much more amenable to experimental measurement and exploitation in future devices.

After the completion of this work\cite{CWG19}, a related manuscript has been posted\cite{RAM19}. This work analyzes the role of the Hartree potential as function of band filling, using a real space tight binding description of a twisted graphene bilayer at a magic angle. As far as the two papers overlap, the results are consistent.

{\it Acknowledgments.}
We would like to thank Jos\'e Gonz\'alez and Tobias Stauber for useful conversations.
This work was supported by funding from the European Commission under the Graphene Flagship, contract CNECTICT-604391.
NRW acknowledges support by the UK STFC under grant ST/P004423/1.

\appendix

\section{Numerical calculation of the order parameter}\label{appA}
\begin{figure}
\includegraphics[scale=0.4]{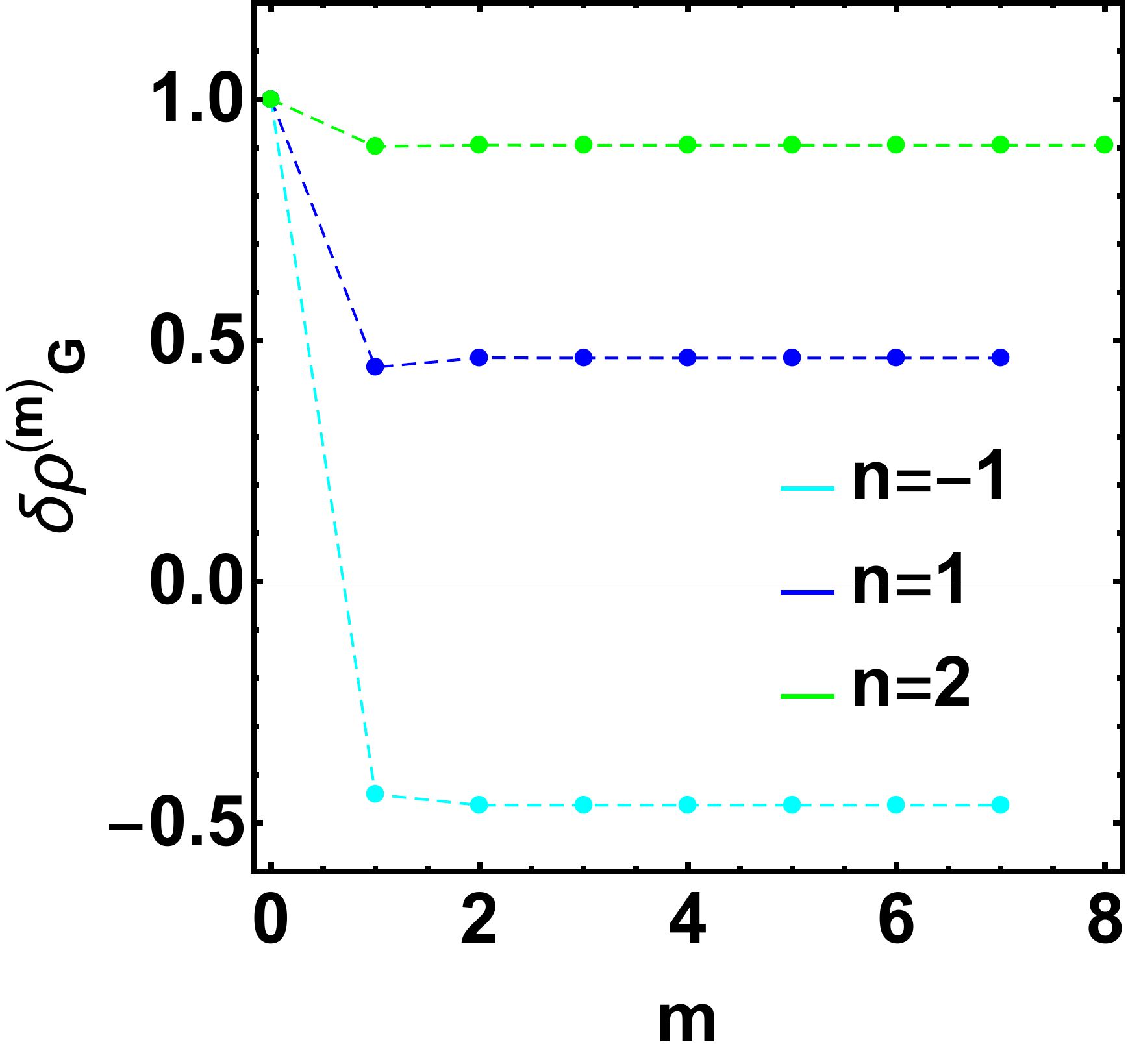}  
\caption{
Value of the order parameter, $\delta\rho^{(m)}_G$, versus the number of steps in the iterative diagonalization, $m$,
for the angle $\theta=1.05^\circ$ and filling $n=-1$ (cyan), $n=1$ (blue) and $n=2$ (green).
}
\label{deltarhoG_vs_step}
\end{figure}

As we already mentioned in the main text, the order parameter $\delta\rho_G$, which defines the Hartree potential, Eq. ~\pref{vH2},
is determined as the self-consistent solution of Eq.~\pref{rhoG}.
We compute the value of $\delta\rho_G$ numerically by means of an iterative procedure starting with an initial value, which in general we choose as $1$.
At the $m$-th step of this procedure, we compute the value $\delta\rho^{(m+1)}_G$ by diagonalizing the Hamiltonian of the TBG in the presence of an Hartree potential given by $\delta\rho^{(m)}_G$.
We set the threshold for convergence by imposing $\left|\delta\rho^{(m+1)}_G-\delta\rho^{(m)}_G\right|<10^{-12}$, where $\delta\rho^{(m)}_G$ is dimensionless and generally of $O(1)$.
In the majority of the cases, the algorithm converges in less than 10 steps. As an example, in Fig. \ref{deltarhoG_vs_step} we show the values of $\delta\rho^{(m)}_G$ versus the number of steps, $m$,
for the angle $\theta=1.05^\circ$ and filling $n=-1$ (cyan), $n=1$ (blue) and $n=2$ (green). As can been seen, there are no significant fluctuations of the order parameter while reaching the solution, which ensures the stability of our method.

\section{screening}\label{appB}
\subsection{Screening due to metallic gates.}\label{appendixa}
The screening induced by a metallic gate at distance $d$ from the twisted graphene bilayer can be analyzed using the method of image charges.A single metallic gate changes the Fourier transform of the effective interaction between two point charges in the bilayer. For a gate at distance $D$ from the bilayer, the potential becomes
\begin{align}
    v_\text{scr} ( \vec{ q} ) &= \frac{2 \pi e^2}{| \vec{ q} |} \left( 1 - e^{- | \vec{ q} | D} \right)
\end{align}
where $\vec{ q}$ is a two dimensional vector in the plane of the bilayer. The relevant Fourier components of the Hartree potential are at reciprocal lattice vectors with $| \vec{ G} | = ( 4 \pi \sqrt{3} ) / ( 3 L_M )$ where $L_M$ is the moir\'e lattice length. Hence, the effective interaction is reduced by a factor 
\begin{align}
    v_\text{scr} ( \vec{ G} ) &= v_\text{bare} ( \vec{ G} ) \left( 1 - e^{- \frac{4 \pi \sqrt{3} D}{3 L_M}} \right)
\end{align}
For bilayer gate distances $D \sim L_M$ we obtain that the reduction factor is $1 - e^{(4 \pi )/ \sqrt{3}} \approx 0.9993$.

A second gate on the other side of the bilayer leads to an infinite series of image charges, with alternating signs. The effect of this series can be summed. For simplicity, we assume that the top and bottom gates are at the same distance from the bilayer, $D$, and obtain
\begin{align}
    v_\text{scr} ( \vec{ G} ) &= \frac{2 \pi e^2}{| \vec{ G} |} \frac{ 1 - e^{- | \vec{ G} | D} }{ 1 + e^{- | \vec{ G} | D} }
\end{align}
For $ e^{- | \vec{ G} | D} \ll 1$, the effect of two gates is, approximately, twice the effect of a single gate.

\subsection{Screening due to the upper and lower bands.}\label{appandixb}
The polarization of the bands not considered in the main text does not contribute to the static dielectric constant, $\vec{ q} \rightarrow 0$. The system formed by these bands describes a two dimensional insulator. The dependence of the susceptibility of these bands, $\chi ( \vec{ q} , \omega )$, at $\omega = 0$ and $ | \vec{ q} | \rightarrow 0$ is $\chi ( \vec{ q} , \omega = 0 )  \propto | \vec{ q} |^2 / \Delta$, where $\Delta$ is an energy scale of order of the gap of the insulator. The contribution to the static susceptibility goes as $\delta \epsilon \propto v_{\vec{ q}} \chi ( \vec{ q} , \omega ) = 0 ) \propto | \vec{ q} | \rightarrow 0$. As function of momentum,  $| \vec{ q} | $, the screening of a slab with $N$ layers will be significant for $| \vec{ q} | \gtrsim \Delta / ( N e^2 )$, provided that the total thickness of the slab, $w$, is such that $w \lesssim | \vec{ q} |^{-1}$. Note that this analysis is also valid for thin slabs of hBN layers.

The relevant components of the Hartree potential in twisted bilayer graphene have Fourier components at the lowest reciprocal lattice vectors, $| \vec{ G} | = ( 4 \pi \sqrt{3} ) / ( 3 L )$. We first approximate the screening at this wavevector by considering the screening due to two decoupled graphene layers. The susceptibility of undoped graphene is\cite{WSSG06}
\begin{align}
    \chi ( \vec{ q} ) &= {\cal N}_f \frac{| \vec{ q} |}{16 \hbar v_F}
\end{align}
where ${\cal N}_f = 8$ is the number of Dirac cones. The contribution to the screening function is
\begin{align}
    \delta \epsilon &= v_{\vec{ q}} \chi ( \vec{ q} ) = \frac{\pi e^2}{\hbar v_F}
\end{align}
This result is independent of the wavevector, in our case $\vec{ G}$. For isolated graphene layers, we can take $e^2 / ( \hbar v_F ) \approx 2$, and we obtain $\delta \epsilon \sim 6$. 

A more precise estimate of the internal screening can be obtained from directly studying the continuum model, The screening from the upper bands at low wavevectors, $\vec{ q} \rightarrow 0$, was considered in Ref.~\cite{PRTVW19}. The calculation is simplest in reciprocal space. The density operator $\tilde{\rho}_{\vec{ G}}$ only has finite matrix elements between states with the same momentum in different bands. The static susceptibility can thus be written as
\bea
    \chi ( \vec{ G} )& =& \frac{4}{V_c}\int_{\text{mBZ}}\frac{\,d^2\vec{k}}{V_{\text{mBZ}}} \\{}
    & \times & 2\sum_{\begin{matrix}l\in\text{occ}\\m\in\text{empty}\end{matrix}} \frac{ \left| \sum_{\vec{G'}} \phi^\dagger_{l  \vec{ k}} \left( \vec{ G} + \vec{ G}'\right) \phi_{m  \vec{ k}} \left( \vec{ G}'\right) \right|^2}{\epsilon_{m, \vec{ k}}-\epsilon_{l, \vec{ k}}}\nn
\eea
where labels $l$ and $m$ denote band indices, the factor 4 stands for the spin and valley degeneracy,
and the factor 2 follows from the double counting of the occupied states in the Lindhard function.
Numerical calculations  give $\chi ( \vec{ G} ) \approx \alpha \,  V_c^{-1}$ with $\alpha=0.07\,\mathrm{meV}^{-1}$ where we use the parameters described in the main text. The correction to the screening is $\delta \epsilon_{\vec{ G}} \approx 8$, in reasonable agreement with the previous analytical estimates.

\section{Band structure and pinning in twisted graphene layers with infinitely narrow bands.}

The effects discussed in this work, the change of the band structure of the low energy bands, and the approximate pinning of the van Hove singularities at the Fermi level for certain band fillings, is clearly manifested in a model where the interlayer hopping between the $AA$ lattices is set to zero\cite{SGG12,TKV_cm19}. This model has electron-hole symmetry, a number of almost equally spaced magic angles, where the bandwidth of the lowest bands is strictly zero\cite{SGG12}.The model gives a simpler description of the emergence of magic angles in twisted bilayer graphene, while keeping the most relevant physical features intact.

If we assume that at half filling the Hartree potential is zero throughout the moir\'e unit cell, we can expect that a self consistent Hartree potential will arise away from half filling. This potential induces a finite bandwidth, which is solely due to the electrostatic interaction. As mentioned in the main text, the fact that he number of electronic flavors is 4 makes the exchange term significantly smaller than the Hartree contribution.

\begin{figure}
\includegraphics[width=3.5in]{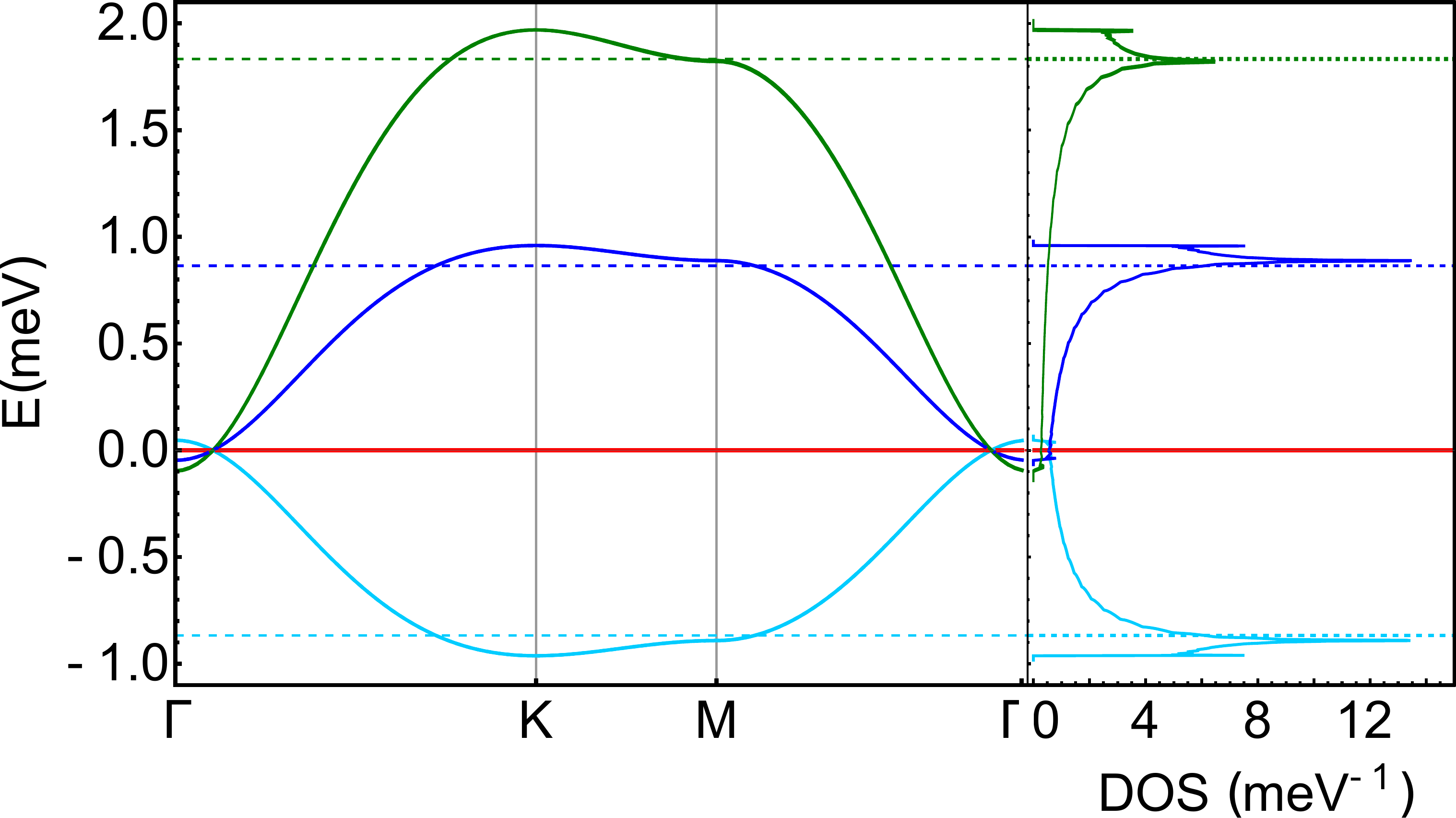}
\caption{
Bands (left) and density of states (right) of the model for twisted bilayer graphene with electron-hole symmetry and infinitely narrow bands at the neutrality point. The fillings considered are $n = -1 , 0 , 1 , 2$, as in Fig.~\ref{bands}. The Fermi velocity, $AB$ interlayer hopping, and relation between the Hartree potential and the charge density used are also as in Fig.~\ref{bands}. The twist angle is $\theta = 1.01^\circ$, which is the first magic angle of the model.
}
\label{fig:chiral}
\end{figure}

Results for the bands and for the density of states are shown in Fig.~\ref{fig:chiral}.
The plots in Fig.~\ref{fig:chiral} indicate that the pinning of the van Hove singularities at integer fillings is enhanced when the bands are infinitely narrow at neutrality. 
The two bands remain degenerate away from neutrality, although they become dispersive. This result can be understood if we assume that the Wannier functions which describe the system at neutrality are the same for the bands away from neutrality. At neutrality, the flatness of the bands implies that all hoppings between Wannier functions are zero. The dispersion shown in Fig.~\ref{fig:chiral} is due solely to new hoppings induced by the Hartree potential. As the model used assumes that the Hartree potential changes in the same way both layers, the two bands acquire a dispersion, but remain degenerate.

\section{Optical absorption.}
A good alternative to STM experiments for studying the van Hove singularities in the density of states experimentally is by observing the optical absorption. In twisted bilayer graphene at the magic angles intra- and interband transitions between the two lowest bands  occur at very low energies, and they may thus be difficult to observe, however.

\begin{figure}
\includegraphics[width=\columnwidth]{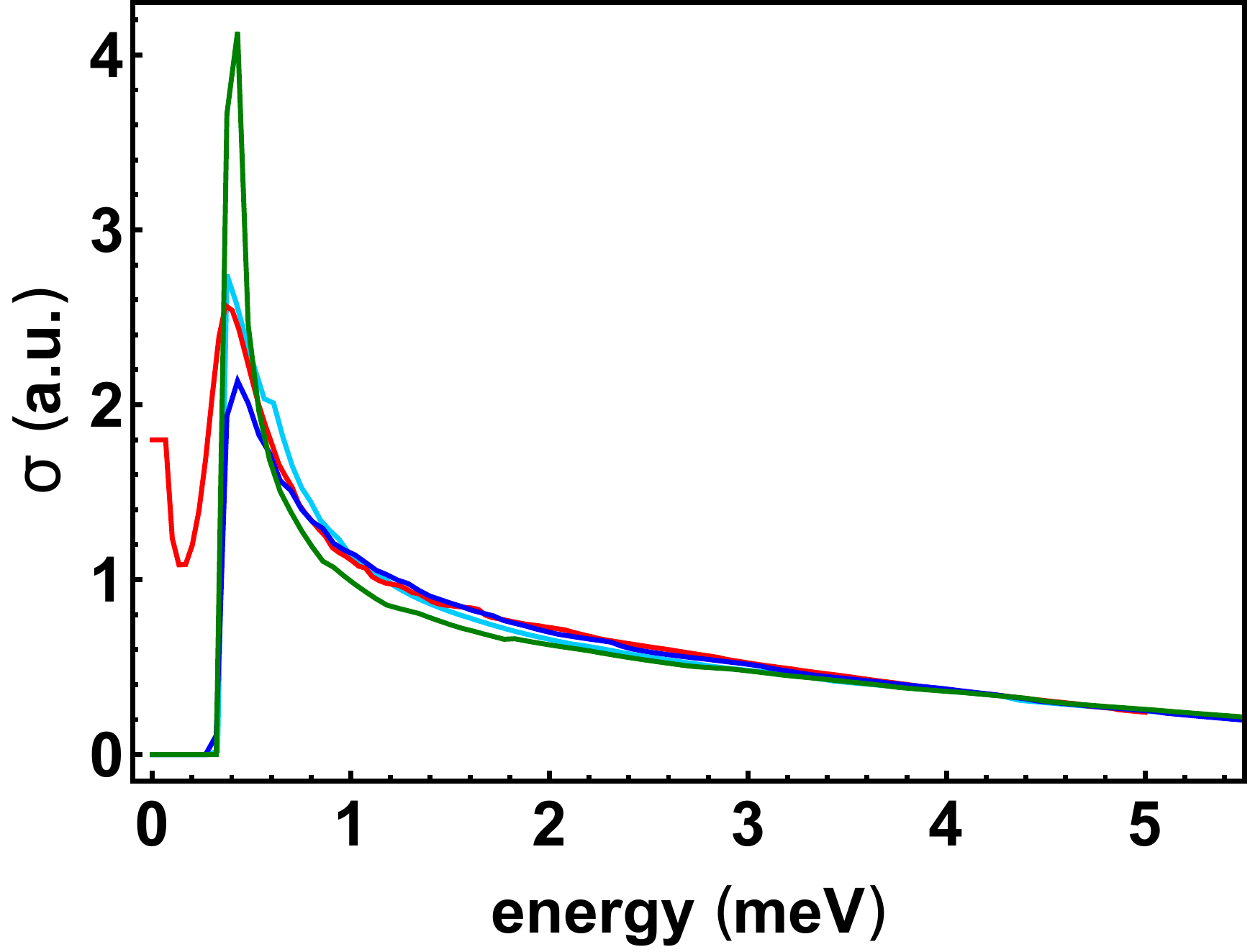}  
\caption{
Optical absorption, in arbitrary units,  for the bands plotted in Fig.~\ref{bands}c. The filligs are: $n = -1, \, {\rm cyan}, n=0, \, {\rm red}, n=1, \, {\rm blue}, n=2, \, {\rm green}$. 
}
\label{fig:optical}
\end{figure}
We have performed a calculation of the optical absorption following the framework set out in \cite{MM13,SSB13}) for the four bands and fillings shown in Fig.~\ref{bands}c. The results are shown in Fig.~\ref{fig:optical}. It is interesting to note that that the energy of the main peak, related to transitions between van Hove singularities, hardly changes with filling. The van Hove singularities in the density of  states associated to the peaks in the optical absorption are in different bands for $n=0$, and in the same bands for $n \ne 0$. Notice that the main peak occurs at $0.5\,\text{meV}$, which is of course a hard wavelength to observe.
\newpage
\bibliography{Literature}
\end{document}